# Enhancement of exchange bias in ferromagnetic/antiferromagnetic core-shell nanoparticles through ferromagnetic domain wall formation

*Rui Wu [1,3,4]\*, Shilei Ding [1,3], Youfang Lai [1,3], Guang Tian [1,3], Jinbo Yang [1,2,3]\**

[1] State Key Laboratory for Mesoscopic Physics, School of Physics, Peking University, Beijing 100871, P. R. China

[2] Collaborative Innovation Center of Quantum Matter, Beijing 100871, P. R. China

[3] Beijing Key Laboratory for Magnetoelectric Materials and Devices，Beijing 100871, P. R. China

4. Department of Materials Science and Metallurgy, University of Cambridge, Cambridge, CB3 0FS, United Kingdom

**Abstract:** The spin configuration in the ferromagnetic part during the magnetization reversal plays a crucial role in the exchange bias effect. Through Monte Carlo simulation, the exchange bias effect in ferromagnetic-antiferromagnetic core-shell nanoparticles is investigated. Magnetization reversals in the ferromagnetic core were controlled between the coherent rotation and the domain wall motion by modulating ferromagnetic domain wall width with parameters of uniaxial anisotropy constant and exchange coupling strength. An anomalous monotonic dependence of exchange bias on the uniaxial anisotropy constant is found in systems with small exchange coupling, showing an obvious violation of classic Meiklejohn-Bean model, while domain walls are found to form close to the interface and propagate in the ferromagnetic core with larger uniaxial anisotropy in both branches of the hysteresis. The asymmetric magnetization reversal with the formation of a spherical domain wall dramatically reduces the coercive field in the ascending branch, leading to the enhancement of the exchange bias. The results provide another degree of freedom to optimize the magnetic properties of magnetic nanoparticles for applications.

**Introduction**

Exchange bias has been found in magnetic materials containing exchange-coupled interfaces between two different magnetic phases. Being of great interest both for the applications in spintronic devices and fundamental physics of condensed matter physics, exchange bias has been extensively studied since its first discovering more than half a century ago[1]. Exchange bias has been found in a wide range of materials including low dimensional composites with ferri-/ferromagnet (FM)/antiferromagnet (AFM) combinations[2-4], single-phase bulk materials with spin glass (SG) or super spin glass (SSG)[5-7]. Recently, giant exchange bias effect has also been reported in magnetic single phases with inter-sublattice interactions[8-10]. Typical systems such as FM/AFM bilayer, and FM/AFM core-shell nanoparticles usually serves as prototype models for the study of exchange bias, due to the well-defined FM and AFM parts as well as the interfaces. Magnetic nanoparticles are of increasing appeal and have found numerous applications in engineering (magnetic recording media or magnetic seals) and biomedical applications (magnetic resonance imaging, drug delivery, or thermotherapy)[11]. With the first exchange bias reported in Co/CoO core-shell nanoparticles, in recent years, the study of exchange bias in nanoparticles and nanostructures has gained renewed interest since it has been shown that control of the core/shell interactions or of the exchange coupling between the particle surface and the embedding matrix can increase the superparamagnetic limit for their use as magnetic recording media[12]. Advances in techniques for synthesis of nanomaterials[13-15] allow the magnetic properties in both the core and the shell to be continuously controlled with morphology[16-19] and composition[20-23] tailoring. A number of factors have shown, in both experimental studies and Monte Carlo (MC) simulations, strong effects in the observed exchange bias or magnetic properties in the core-shell structures, including the core/shell thicknesses[24-27], particle shape/morphology[28,29], cooling field[30,31], dipolar interactions[32,33]/interparticle exchange interactions[11], and interface lattice/magnetic disorder/mismatch[34,35].

In the Meiklejohn-Bean (M-B) model, by assuming a collinear magnetization reversal in both FM and uncompensated AFM parts, the exchange bias field was predicted to be

$$h_E = \frac{\sigma_{ex}}{t_{FM} M_{FM}} \quad (1)$$

where $\sigma_{ex}$, $t_{FM}$, and $M_{FM}$ stand for the interfacial exchange coupling energy, the FM thickness, and the FM magnetization, respectively. Thus, an inversely linear dependence on the thickness of FM layer[36,37] and no dependence on intrinsic properties of FM part, including the magnetic anisotropy and the exchange coupling strength were indicated in the model. Since the M-B model works very well in many systems, the effect of inner magnetic structure in the FM part has been overlooked to some extent for quite a long time while the most effort has been devoted to the magnetic structures in AFM parts and interfaces. However, recent experimental and theoretical results indicate that this rule can be violated while a partial domain wall parallel to FM-AFM interface forms in FM layer during the magnetization reversal process[38,39]. Although nonuniform magnetization configurations have been reported in magnetic nanoparticles *via* small-angle neutron scattering[40,41] magnetic force microscopy[42], magnetic electron holography[43] and MC simulations[44-47], its effect on the exchange bias of FM-AFM core-shell structures and how it can be controlled remain unknown.

In this paper, it is shown, through MC simulations based on a simple model of single core/shell nanoparticle, how the formation of a spherical domain wall in the FM core is related to exchange bias in this system. The spherical domain wall is induced or suppressed in the FM core by tuning the domain wall width by varying the anisotropy constant and the exchange coupling strength. This result is confirmed by inspection of magnetic configurations and curls of magnetic configurations in the core along the hysteresis loops. It is further demonstrated that the formation of a spherical domain wall in the core while magnetization reversal significantly reduces the coercive field in the ascending branch, and consequently enhances the exchange bias field.

**Model**

The considered nanoparticles have a spherical shape with a total radius of $R = 12a$, respectively, with $a$ being the unit cell size. All the particles are made of an FM core surrounded by an AF shell of a constant thickness $R_{Sh} = 3a$ with magnetic properties different from the core as well as from the spins at the interface between core and shell spins. Taking $a = 0.3$ *nm*, such a particle corresponds to typical real dimensions R $\approx$ 4 *nm* with a fixed shell thickness of $R_{Sh}$ $\approx$ 1 *nm* and contains 5575 spins, with 3071 spins in the FM core and 2504 spins in the AFM shell. The interface is defined to be the atoms in the AFM shell which have direct exchange coupling with the FM core and contains 918 spins. The anisotropic Heisenberg spin model is adopted in the calculations with a Hamiltonian given by

$$H/k_B = -J_{FM} \sum_{\langle i,j \in FM \rangle} \vec{S}_i \cdot \vec{S}_j - J_{AFM} \sum_{\langle i,j \in AFM \rangle} \vec{S}_i \cdot \vec{S}_j - J_{INT} \sum_{\langle i \in FM, j \in AFM \rangle} \vec{S}_i \cdot \vec{S}_j \\ - K_{FM} \sum_{\langle i \in FM \rangle} \vec{S}_{iz}^2 - K_{AFM} \sum_{\langle i \in AFM \rangle} \vec{S}_{iz}^2 \\ - \sum_{i=1}^{N} \vec{h} \cdot \vec{S}_i \quad (2)$$

where $\vec{S}_i$ are classical Heisenberg spins of unit magnitude placed at the nodes of a simple cubic lattice. The first row gives the exchange energy between spins located in FM core, AFM shell and FM-AFM interface with exchange coupling constants denoted by $J_{FM}$, $J_{AFM}$ and $J_{INT}$, respectively. The second row gives the local anisotropic energy for each spin in FM core and AFM shell with the anisotropy constant represented by $K_{FM}$ and $K_{AFM}$, respectively. The local anisotropy axes are set to be the *z*-direction for all spins to impart a uniaxial anisotropy to the simulated systems. The last term describes the Zeeman coupling to an external field $H$ applied along the easy-axis direction, which in reduced units reads $\vec{h} = \mu \vec{H}/k_B$ (with $\mu$ the magnetic moment of the spin) and will be denoted in temperature units[48].

To calculate the magnetic properties, the MC method with a standard Metropolis algorithm is employed[49]. As for the spin updates, an attempt to change the spin at a randomly picked site *i* from $\vec{S}_i$ to $\vec{S}'_i$ is made in a Monte Carlo trial step with the

acceptance rate given by

$$P(\vec{S}_i \rightarrow \vec{S}_i') = \min\left[1, \exp\left(-\Delta E / k_B T\right)\right] \quad (3)$$

where $\Delta E$ denotes the change in free energy of the system if $\vec{S}_i'$ is accepted. To get an optimum efficiency for the Heisenberg system with finite uniaxial anisotropies, a combination of three kinds of trial steps, a uniform movement, a small movement, and a reflection, with a ratio of 3:1:1, is adopted[50]. In the uniform movement, the direction of $\vec{S}_i'$ is selected by random sampling on a sphere with Marsaglia method[51]. In the small movement, the direction of $\vec{S}_i'$ is selected by random sampling in a cone centered about $\vec{S}_i$. A reflection movement, where the direction of $\vec{S}_i'$ is selected to be $-\vec{S}_i$, is included to simulate nucleation processes even more efficiently in the limit of very large anisotropy.

An MC step (MCS) is finished while every spin in the whole system has undergone a trial step for once. To get the equilibrium state, at each field (or temperature) point, 10000 MCSs are performed with 9800 MCSs for configuration relaxation and the remaining 200 MCSs for averaging the quantities, which is enough to minimize the fluctuation in the data, especially at low temperature. To get more detailed magnetization reversals around the coercive field, smaller field steps are used for the spin configuration calculation with keeping the total MCSs.

**Results and discussion**

1. **The $K_{FM}$ dependence of exchange bias**

Systems with different ferromagnetic anisotropy constants, $K_{FM}$, are field cooled (FC) from a high-temperature (far above Néel temperature of AFM shell, $T_N$) disordered phase in a constant step down to the measuring temperature $T = 0.1~K$ in the presence of a cooling field $h_{FC} = 0.4J_0$ applied along the easy-axis direction, with $J_0 = 10~K$ as a reference parameter. All the other parameters, the exchange coupling in FM core $J_{FM} = J_0$, exchange coupling at the FM-AFM interface $J_{INT} = -0.5J_0$, exchange coupling in AFM shell $J_{AFM} = -0.5J_0$, anisotropy constant in AFM shell $K_{AFM} = J_0$, were kept the same within all systems, which was targeted to give a larger Curie temperature $T_C$ of FM core than $T_N$ and a relatively large anisotropy of AFM part due to the ultrathin thickness of the AFM shell[28]. The $K_{FM}/J_0$ is varied from 0 to 0.1, which is in the reasonable range for real ferromagnetic systems[52]. The temperature dependence of the normalized magnetizations $M/M_S$ (with $M_S$ being the total number of spins in the nanoparticle) in core, shell, and interface in a system with $K_{FM}/J_0 = 0.1$ is given in FIG. 1(a), where a paramagnetic to ferromagnetic transition is observed when temperature decreases across the Curie temperature ($T_C \approx 15~K$) of the FM core and a paramagnetic to antiferromagnetic transition is observed when temperature decreases across the Néel temperature ($T_N \approx 6.5~K$) of the AFM shell. Due to the antiferromagnetic exchange coupling at the interface between the FM core and the AFM shell, the uncompensated interfacial spins give a negative net magnetization. From FIG. 1(b), the interfacial net magnetization $M_{INT}$ remains nearly invariant with increasing $K_{FM}$ at all temperatures, indicating that the spin configuration in the FM core is dominated by the exchange coupling $J_{FM}$ and the cooling field $h_{FC}$.

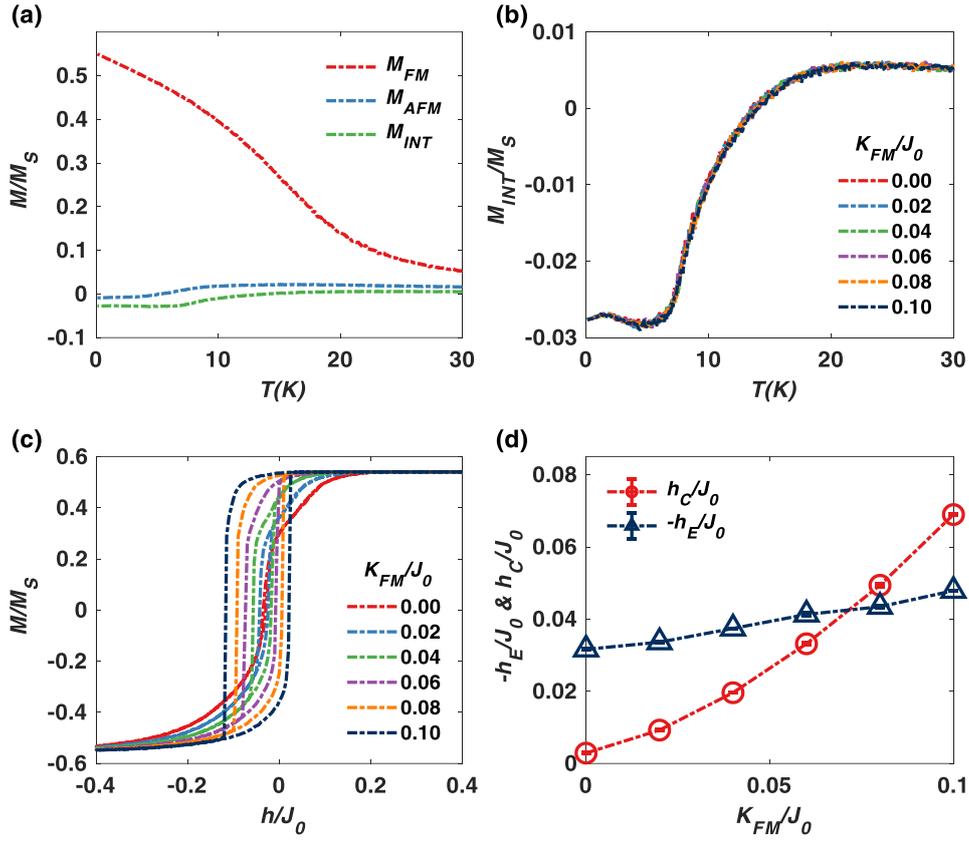

**FIG. 1.** (a) The FC *M-T* curves of different parts of the core-shell structure with $K_{FM}/J_0 = 0.1$. (b) The FC *M-T* curves of interfacial spins, (c) hysteresis loops after FC, and (d) extracted $h_E$ and $h_C$ obtained in core-shell structures with $0 \leqslant K_{FM}/J_0 \leqslant 0.1$. All the data in (d) are averaged with three independent calculations with error bars coming from the calculated standard deviations.

After the FC, hysteresis loop calculations are undertaken for each system with different $K_{FM}$ using the starting configuration obtained with the FC process and by cycling the magnetic field from $h = 0.4J_0$ to $h = -0.4J_0$ in steps $h = -0.005J_0$. Integration of the magnetization is carried out over the whole system. As shown in FIG. 1(c), the hysteresis loops change significantly with the increasing uniaxial anisotropy constant of the FM core. As expected, a larger $K_{FM}$ unambiguously gives a larger coercivity in the hysteresis loop where nearly zero coercive fields were obtained with $K_{FM} = 0$ with the hard-axis switching characteristics presented, showing a progressive approach to both positive and negative saturation, due to the spin-flop coupling between FM spins and those compensated AFM spins at the interface[53,54,55]. As the $K_{FM}$ increases, the induced anisotropy perpendicular to $z$-axis is overwhelmed by the uniaxial anisotropy of the FM core itself, showing a sharper magnetization switching in both sides and a significantly enhanced coercivity. However, as shown FIG. 1(d), it is found that the dependence of the coercivity $h_C$ [defined as $h_C = (h_{CR}-h_{CL})/2$ where $h_{CR}$ and $h_{CL}$ are the left coercive field and right coercive field, respectively] on the $K_{FM}$ is not linear. Moreover, the exchange bias fields $h_E$ [defined as $h_E = (h_{CR}+h_{CL})/2$] also shows a monotonic increase with increasing $K_{FM}$ which violates the result predicted by M-B model where the exchange bias field only depends on the interfacial exchange coupling energy $\sigma_{ex} \sim J_{INT}M_{INT}$ and the total magnetization of the FM part $t_{FM}M_{FM}$. Since both $J_{INT}$ and $t_{FM}$ are invariant with $K_{FM}$, to reveal the underlying origin of this effect, constrained MC calculations are undertaken, in which the AFM spins are fixed in the hysteresis loop calculations after the same FC process with the non-constrained MC calculations. Thus, the effect of the FM core behavior on the exchange bias can be studied separately.

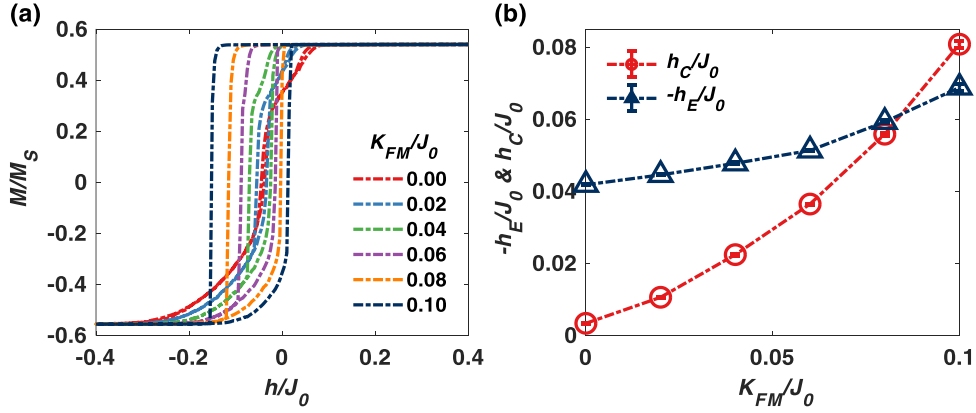

**FIG. 2.** (a) The hysteresis loops calculated with constrained MC and (b) the extracted $h_E$, $h_C$ in core-shell structures with $0 \leqslant K_{FM}/J_0 \leqslant 0.1$ after FC. All the data in (b) are averaged with three independent calculations with error bars coming from the calculated standard deviations.

As shown in FIG. 2, the hysteresis loops calculated with the constrained MC show similar $K_{FM}$ dependence with those obtained with non-constrained MC method especially when the $K_{FM}$ is small, where the hysteresis also shows a hard-axis like magnetization switching coming from the spin-flop coupling. However, with higher $K_{FM}$, the hysteresis shows higher asymmetry with sharper magnetization switching in the descending branch than the one obtained with non-constrained MC; this effect being ascribed to the rigidness of the interfacial AFM spins in the constrained MC. Meanwhile, both $h_C$ and $h_E$ given in FIG. 2(b) are larger than those obtained with non-constrained MC, indicating stronger pinning effect of the constrained AFM magnetic moments. Further, it is worth noting that the $K_{FM}$ dependence of $h_E$ and $h_C$ shows similar behavior with those obtained from non-constrained MC calculation with monotonic dependence with $K_{FM}$. An increment of 64.6% in $h_E$ is obtained in the hysteresis loop with $K_{FM}/J_0 =$ 0.1 compared to that with $K_{FM}/J_0 = 0$, which is even a little larger than the result 51.4% obtained in non-constrained MC.

Since the AFM spins are fixed in the constrained MC, it is demonstrated that the monotonic increase of $h_E$ and the non-linear increase of $h_C$ with the increasing $K_{FM}$ is contributed by the FM core. This can be corroborated by direct inspection of the spin configurations along the loops, as presented in the main panel of Fig. 3 for $K_{FM}/J_0 = 0$. As it is evidenced by the sequence of snapshots, the reversal proceeds by quasi-uniform rotation along both descending and ascending branches at magnetic fields around left and right coercive fields, respectively. The hysteresis loop shows different approaching behaviors to the two saturation directions, although both are reversible. The progressive approaching to negative saturation has been proven to originate from a planar domain wall formed parallel to the FM/AFM interface[56,57]. As shown in FIG. 3(c) and (d), this domain wall is also observed with a spherical shape in the core-shell nanoparticle where the spins close to core center reverse before those close to the core-shell interface in the descending branch (FIG. 3(c)). While in the approaching to positive saturation (FIG. 3(a) and (f)), all the spins in the core rotate coherently without formation of the domain wall.

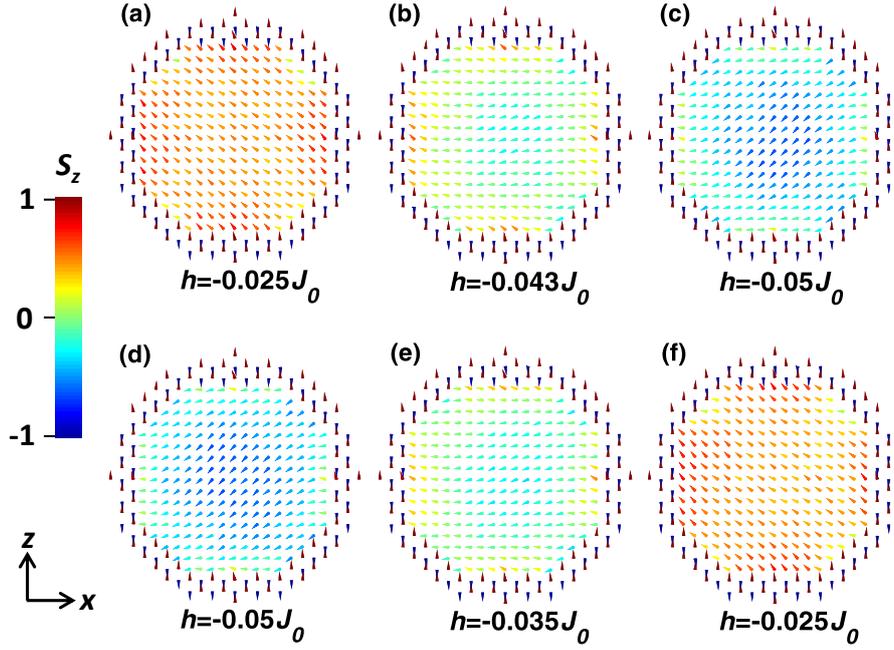

FIG. 3. Snapshots of spin configurations during magnetization reversals around the left coercive field (a-c) and right coercive field (d-f) in the system with $J_{FM}/J_0$ = 1 and $K_{FM}/J_0$ = 0, calculated with constrained MC. The color of the arrow indicates the magnitude of the z component of each spin.

For comparison, spin configurations of the nanoparticle with FM anisotropic constant of $K_{FM}/J_0 = 0.1$ are inspected. As shown in FIG. 4, the magnetization reversal along the descending branch proceeds first with quasi-uniform rotation and then with a fast propagation of planar domain wall nucleated at one point of the interface, while the nucleation of reversed domains at the whole interface and its subsequent slow shrink across the core center is the major reversal process along the ascending branch, resulting in an asymmetric characteristic in the hysteresis loop. Similar asymmetry in hysteresis loops also has been observed experimentally in discontinuous nanostructure[41,2]. The asymmetric magnetization reversals here have similar features but different mechanisms from those obtained in continuous films, where the domain wall motion occurs in the descending branch while domain rotation occurs in ascending branch[58,59], originating from a biaxial magnetic anisotropy in the AFM part[60].

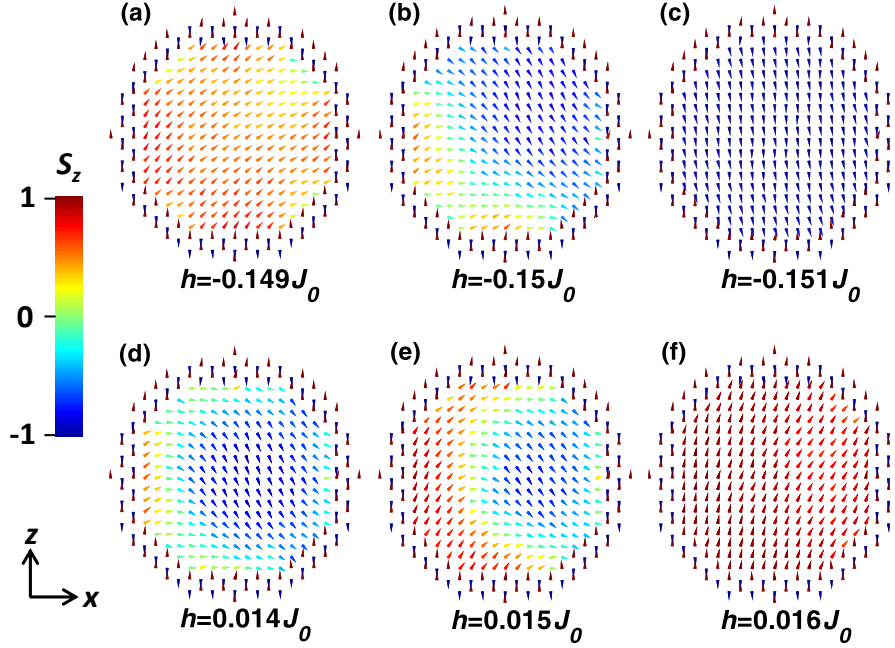

**FIG. 4. Snapshots of spin configurations during magnetization reversals around the left coercive field (a-c) and right coercive field (d-f) in the system with $J_{FM}/J_0$ = 1 and $K_{FM}/J_0$ = 0.1, calculated with constrained MC. The color of the arrow indicates the magnitude of the z component of each spin.**

Here, it is demonstrated that the exchange bias field is strongly correlated with special reversal mechanism in the core-shell nanoparticle. First, $\vec{R}$, the curl of the spin vector field $\vec{S}(S_x, S_y, S_z)$, is used to describe the non-collinearity of spin configuration in the FM core, which reads

$$\vec{R} = \nabla \times \vec{S} = R_x \vec{i} + R_y \vec{j} + R_z \vec{k}$$
$$= \left( \frac{\partial S_z}{\partial y} - \frac{\partial S_y}{\partial z} \right) \vec{i} + \left( \frac{\partial S_x}{\partial z} - \frac{\partial S_z}{\partial x} \right) \vec{j} + \left( \frac{\partial S_y}{\partial x} - \frac{\partial S_x}{\partial y} \right) \vec{k} \quad (4)$$

with $R_x$, $R_y$ and $R_z$ representing three components of local curls. The differentials are calculated with finite difference method. The vortex-like local spin configurations will yield non-zero local curls while the collinear spin configurations will give zero local curls. The overall magnitude of the microscopic (local) curls can be given as

$$C_{micro} = \sum_{i \in FM} \sqrt{R_{ix}^2 + R_{iy}^2 + R_{iz}^2} \quad (5)$$

which enable us to get an insight into the noncollinearity of FM core while the magnitude of macroscopic (global) curls can be represented as

$$C_{macro} = \sqrt{\left( \sum_{i \in FM} R_{ix} \right)^2 + \left( \sum_{i \in FM} R_{iy} \right)^2 + \left( \sum_{i \in FM} R_{iz} \right)^2} \quad (6)$$
$$= \sqrt{C_x^2 + C_y^2 + C_z^2}$$

which enables us to investigate the evolution of macroscopic curling while the orientation of macroscopic curling can be obtained with its components $C_x$, $C_y$ and $C_z$.

As shown in FIG. 5, within a core with $K_{FM}/J_0 = 0$, both overall microscopic curling $C_{micro}$ and macroscopic curling $C_{macro}$ show very small deviations at all fields from saturation states, which confirms magnetization reversals in the core are nearly coherent in both branches. However, it is worth noting that there is a significant shoulder at the left side of each coercive field in $C_{micro}$, which is absent in $C_{macro}$, while both $C_{micro}$ and $C_{macro}$ show two peaks at coercive fields. From spin configurations given in FIG. 3 (c) and (d), the shoulders in $C_{micro}$ are related to the formation of spherical domain walls in these field regions. In the spherical domain wall, the local curl at one point is opposite to that at its symmetric point, giving zero contribution to the macroscopic curl. Thus, in macroscopic curls, two peaks without shoulders around coercive fields are obtained, which are also present in microscopic curls. As $K_{FM}/J_0$ increases, the right peak shows a monotonic increase while the left peak nearly does not change. From FIG. 4 (b) and (f), it can be inferred that peaks at the left coercive fields and the right coercive fields are related to a planar domain wall and an incomplete spherical domain wall, respectively. The peak shoulder in $C_{micro}$, which is related to a complete spherical domain wall, maintains in ascending branches but decreases in descending branches and finally disappears in the system with $K_{FM}/J_0 = 0.1$, showing asymmetric magnetization reversal in the two branches. Three components of macroscopic curls show similar dependence to $K_{FM}/J_0$. $C_x$ and $C_y$ always follow each other due to the rotation symmetry of the considered systems in the $x$-$y$ plane. With increasing $K_{FM}/J_0$, both $C_x$ and $C_y$ peaks increase monotonically in descending branches but keep nearly invariant in ascending branches. Differently, $C_z$ only occurs in ascending branches in systems with $K_{FM}/J_0 > 0.06$, indicating the emerging of a curling in the $x$-$y$ plane.

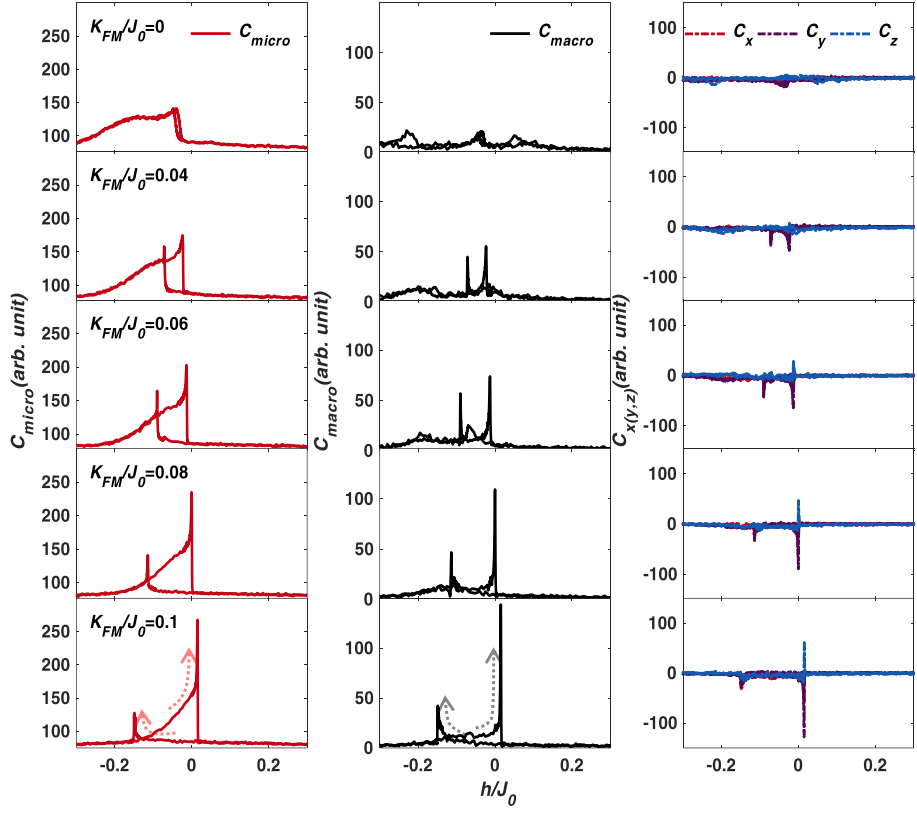

**FIG. 5.** The overall microscopic curls (left column), the macroscopic curls (middle column) and three different components of macroscopic curls (right column) of systems with $J_{FM}/J_0 = 1$ and $0 \leqslant K_{FM}/J_0 \leqslant 0.1$.

## 2. Effect of exchange strength $J_{FM}$

The anomalous dependence of $h_E$ and $h_C$ on $K_{FM}$ may originate from this asymmetric magnetization reversal behavior, which is related to different domain structures formed in descending and ascending branches. The formation of a spherical domain wall in the ascending branch of the hysteresis loop can effectively reduce the increment of right coercive field caused by increasing $K_{FM}$ and consequently increases $h_E$ (as shown in FIG. 2). For a classic approximation, the domain wall width of a ferromagnetic material is determined by the competition between exchange and effective anisotropic energy,

which is given by

$$\delta_w = \pi\sqrt{\frac{A}{K}} = \pi\sqrt{\frac{J_{FM}}{K_{FM}}} \quad (7)$$

where $A = nS^2 J_{FM}/a = J_{FM}$ (with $n=1$ for simple cubic structure, $S = 1$, $a = 1$ for considered systems) is the exchange stiffness constant and $K = K_{FM}$ is the anisotropy constant of the material.

The $K_{FM}$ and $J_{FM}$ dependences of $\delta_w$ are plotted in FIG. 6. For a given $J_{FM}$, $\delta_w$ decreases sharply at the beginning and then gradually in the end with the increasing $K_{FM}$. For a given $K_{FM}$, a smaller $\delta_w$ is obtained with a small $J_{FM}$ than that obtained with a large $J_{FM}$. Consequently, given a smaller $J_{FM}$ and a larger $K_{FM}$, the $\delta_w$ will be small enough to enable domain wall formation in the FM core with a diameter of $18a$. Also, the domain wall in FM core will be suppressed with larger $J_{FM}$ and smaller $K_{FM}$.

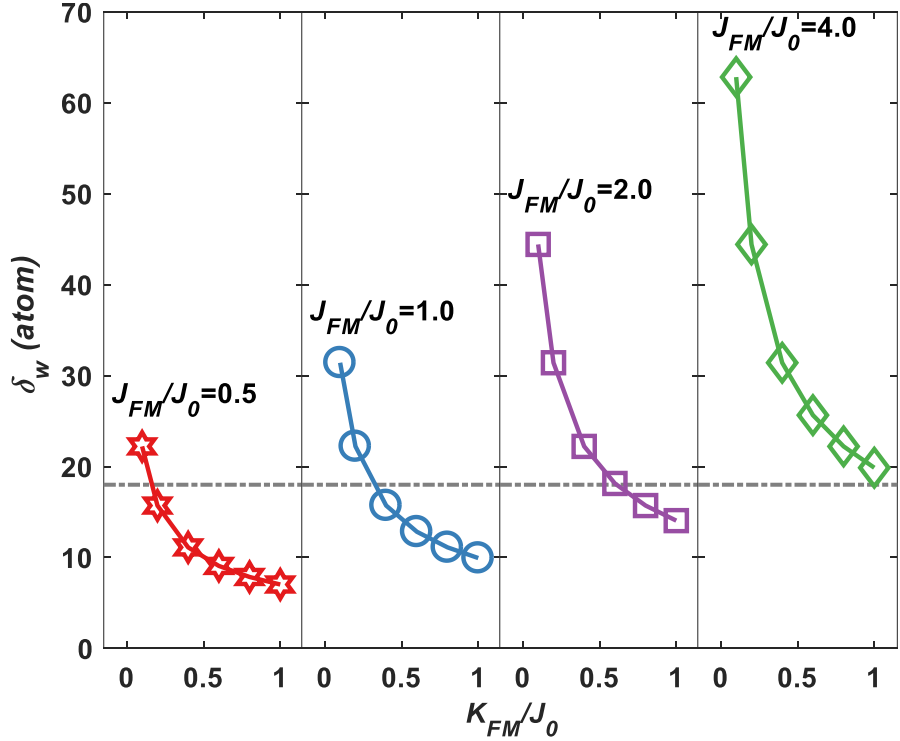

**FIG. 6. The dependence of domain wall width on $K_{FM}$ and $J_{FM}$ calculated from Eqn. (7) with a grey dashed line indicating the diameter of the FM core.**

To verify this hypothesis, $K_{FM}$ dependence of exchange bias in this system with varying $J_{FM}$ is studied. In all the calculations, AFM spins are constrained. As shown in FIG. 7(a), the exchange bias field $h_E$ shows a very sharp increase with increasing $K_{FM}$ with a small ferromagnetic exchange coupling $J_{FM}/J_0 = 0.5$. This monotonic dependence of $h_E$ on $K_{FM}$ maintains with increasing $J_{FM}$ up to $J_{FM}/J_0 = 2$ and finally disappears in the system with $J_{FM}/J_0 = 4$, where $h_E$ shows no obvious dependence on $K_{FM}$. The effect of $J_{FM}$ is more prominent in the relative increment of exchange bias field, $\delta h_E/h_{E0}$, where $h_{E0}$ and $\delta h_E$ are the $h_E$ at $K_{FM}/J_0 = 0$ and the increment of $h_E$ relative to $h_{E0}$ at $K_{FM}/J_0 \neq 0$. As shown in FIG. 7(c), $\delta h_E/h_{E0}$ shows a very sharp increase with increasing $K_{FM}$ in a system with $J_{FM}/J_0 = 0.5$. The increase is largely reduced in systems with larger $J_{FM}$. Finally, a nearly zero increment in $h_E$ is obtained with increasing $K_{FM}$ in the system with $J_{FM}/J_0 = 4$.

Meanwhile, $h_C$ also shows a strong dependence on both $K_{FM}$ and $J_{FM}$. As shown in FIG. 7(b), with a small $J_{FM}$, the system shows superparamagnetic characteristic with nearly zero $h_{C0}$ (the $h_C$ at $K_{FM}/J_0 = 0$). As the $J_{FM}$ increases, $h_{C0}$ shows a monotonic increase, indicating an increasing magnetic anisotropy given by the exchange coupling at the core-shell interface. Consequently, the relative increment of coercivity, $\delta h_C/h_{C0}$, where $\delta h_C$ is the increment of $h_C$ relative to $h_{C0}$ at $K_{FM}/J_0 \neq 0$, increases with $K_{FM}$ but decreases with $J_{FM}$, as shown in FIG. 7(d). Moreover, it is found that the way in which $h_C$ depends on $K_{FM}$ varies with $J_{FM}$ significantly. When $J_{FM}$ is small, $h_C$ shows a nonlinear dependence on increasing $K_{FM}$, with a gradual increase at lower $K_{FM}$ and a steeper increase at higher $K_{FM}$. However, when $J_{FM}$ increases, the nonlinearity of the dependence is reduced and, finally, becomes a linear dependence in the system with $J_{FM}/J_0 = 4$.

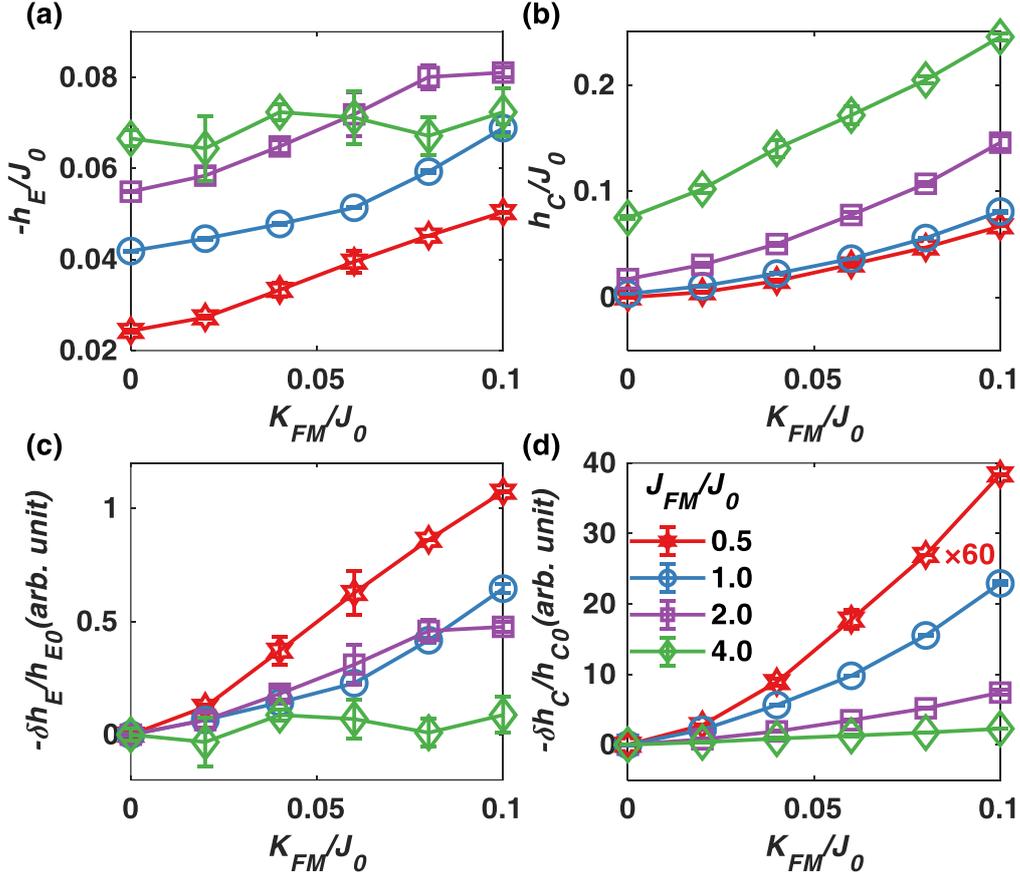

**FIG. 7.** The $K_{FM}$ dependence of (a) $h_E$, (b) $h_C$, (c) relative change of $h_E$ and (d) relative change of $h_C$ plots with different $J_{FM}$. All the data are averaged with three independent calculations with error bars coming from the calculated standard deviations.

An invariant $h_E$ and a linear dependent $h_C$ on $K_{FM}$ is exactly the results predicted by M-B single spin model, which is absent in system with small $J_{FM}$ and presents in system with large $J_{FM}$, indicating an evolution of the spin configuration from non-collinear to collinear during magnetization reversals as $J_{FM}$ increases, which is verified by an inspection of spin configurations and overall microscopic curls of the systems with different $J_{FM}$ during magnetization reversals.

It can be seen from the first column of FIG. 8, the planar domain wall at the left coercive field shows strong dependence with the increasing $J_{FM}$. The planar domain wall with a small width is very significant in a system with small $J_{FM}$, and becomes weaker with a larger domain wall width as $J_{FM}$ increases. A collinear alignment of core spins and decreased contrast in color map of $S_z$ are observed in the system with $J_{FM}/J_0 = 4.0$, as shown in FIG. 8(j). The spherical domain wall at the right coercive field shows similar $J_{FM}$ dependence as the planar domain wall, as shown in middle column of FIG. 8, which becomes weaker and broader with increasing $J_{FM}$ and nearly disappears in the system with $J_{FM}/J_0 = 4.0$. The evolution of domain structure with $J_{FM}$ is also reflected in overall microscopic curls (FIG. 8, right column). As shown in FIG. 8 (c), overall microscopic curls in the system with $J_{FM}/J_0 = 0.5$ are very large with contributions including a large background coming from the random thermal fluctuation, two peaks from planar domain wall and incomplete spherical domain wall at left coercive field and right coercive field, respectively, and broad shoulders from the complete spherical domain walls. With an increased $J_{FM}/J_0$, as shown in FIG. 8(f) and 8(i), overall microscopic curls are lowered significantly, which is in good agreement with the spin configurations. Meanwhile, the background is also reduced largely, which is ascribed to the effectively suppressed thermal fluctuations by the large exchange coupling. Finally, as shown in FIG. 8(l), with the largest $J_{FM}/J_0$ of 4.0, both the peaks and the background are largely reduced corresponding to nearly collinear spin configurations during the magnetization reversals.

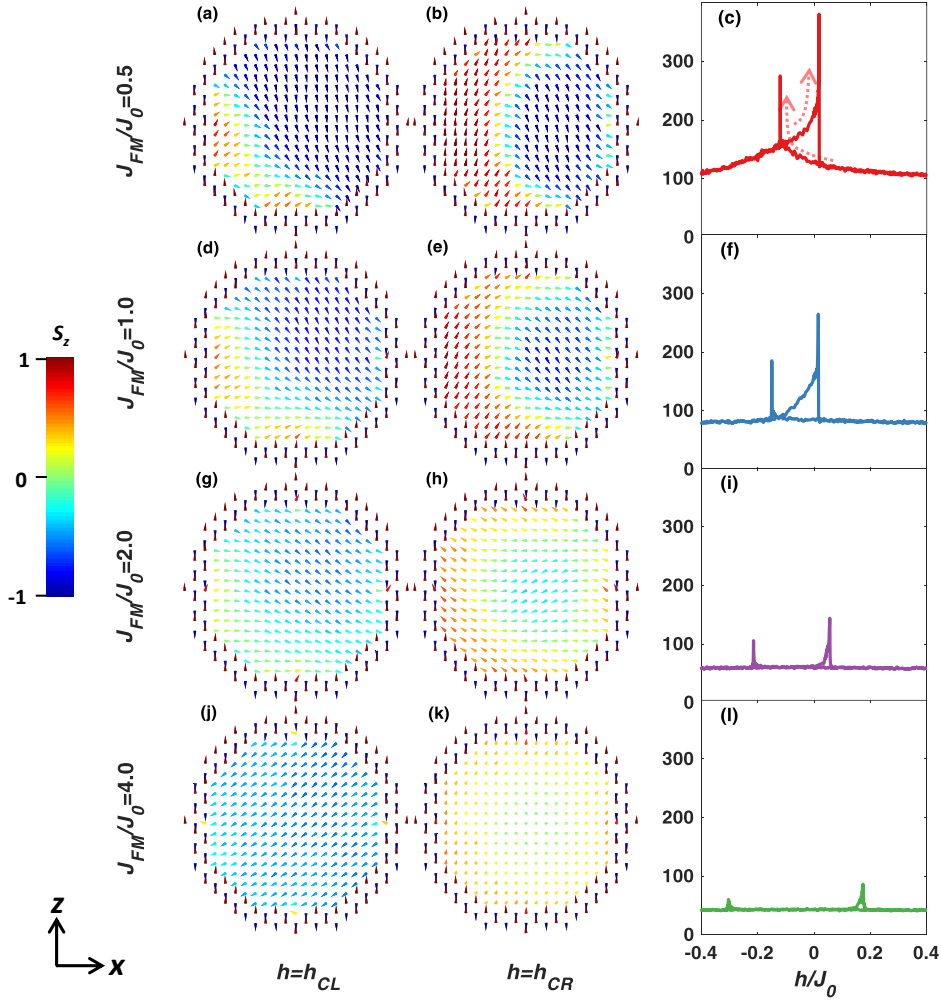

**FIG. 8. Spin configuration snapshots of systems with the same $K_{FM}/J_0 = 0.1$ but with $0.5 \leqslant J_{FM}/J_0 \leqslant 4.0$ taken at left coercive fields (left column) and right coercive fields (middle column) and overall microscopic curls as functions of the magnetic field in systems with different $J_{FM}$, calculated with constrained MC. The color of the arrows in the spin configuration snapshots indicates the magnitude of the z-component of each spin.**

For a realistic material consideration, typical domain wall widths of Fe, Co, and Ni nanoparticles are around 138 *nm*, 36 *nm*, and 285 *nm*, respectively[61]. However, the magnetic vortex state has been observed in Fe nanoparticle with a size of 26 *nm*[43], indicating the noncollinear magnetic configuration can be obtained in magnetic nanoparticles much smaller than the bulk domain wall width. In harder magnetic materials, much smaller domain wall width can be obtained. For instance, domain wall widths for $CoFe_2O_4$ and $Nd_2Fe_{14}B$ are about 8 *nm*[62] and 5 *nm*[63], respectively. An incomplete spherical domain wall can exist in a nanoparticle around this length scale, which can be easily manipulated with size controlling and composition tailoring to give optimized exchange bias effect and other magnetic properties.

**Conclusions**

To conclude, the effect of FM spin configuration on the exchange bias effect of FM/AFM core-shell nanoparticles has been studied with MC method. A significant enhancement of the exchange bias effect accompanied by a nonlinear behavior of coercivity with increasing magnetic anisotropy constant $K_{FM}$ has been observed, showing a violation of classic M-B model. This anomalous effect is ascribed to the asymmetric magnetization reversal in the FM core with a spherical domain wall formation in the ascending branch of the hysteresis loop, which largely reduces the right coercive field and enhances the exchange bias field. This is demonstrated by adjusting the domain wall width in the FM core with varying $J_{FM}$ and $K_{FM}$. Finally, the anomalous dependence of $h_E$ and $h_C$ on $K_{FM}$ disappears when the domain wall in the core is suppressed. The results provide another freedom to tailor the exchange bias in the FM/AFM systems.

**Acknowledgements**

This work is supported by the National Key Research and Development Program of China (No. 2017YFA0206303, 2016YFB0700901 and 2017YFA0401502) and National Natural Science Foundation of China (Grant Nos. 51731001, 51371009, 11504348, 11675006), the Ph.D. Programs Foundation of Ministry of Education of China (No. 20130001110002).

[*] Corresponding author: jbyang@pku.edu.cn and rw556@cam.ac.uk